# A comparison algorithm to check LTSA Layer 1 and SCORM compliance in e-Learning sites

S. Sengupta, S. Pal, N. Banerjee

**Abstract** — The success of e-Learning is largely dependent on the impact of its multimedia aided learning content on the learner over the hyper media. The e-Learning portals with different proportion of multimedia elements have different impact on the learner, as there is lack of standardization. The Learning Technology System Architecture (LTSA) Layer 1 deals with the effect of environment on the learner. From an information technology perspective it specifies learner interaction from the environment to the learner via multimedia content. Sharable Content Object Reference Model (SCROM) is a collection of standards and specifications for content of web-based e-learning and specifies how JavaScript API can be used to integrate content development. In this paper an examination is made on the design features of interactive multimedia components of the learning packages by creating an algorithm which will give a comparative study of multimedia component used by different learning packages. The resultant graph as output helps us to analysis to what extent any LMS compliance LTSA layer 1 and SCORM specification.

## 1 INTRODUCTION

THERE is substantial research work already done to support the effectiveness of multimedia components on e-Learning content development as multimedia can stimulate more than one sense at a time, and in doing so, may be more attention-getting and attention-holding. While the benefits of eLearning are not in dispute, there are a number of impediments to successful use of a e-Learning portal by learners [8]. Recognizing the importance of nurturing individual learning requirements, the design capability should be anchored to established learning and development theories which is applied by e-Learning professionals with proven commercial experience. While there are numerous e-Learning solutions available today, the differentiating factors, i.e., the factors which help achieve a far superior learning experience and motivation to learners, are the innovative instructional design and custom content development process. A good content design should consider the accessibility, clarity, consistency, efficiency, focus, and flexibility of the content. In the context of e-learning technology, standards are generally developed for use in systems design and implementation for the purposes of ensuring interoperability, portability and reusability. These attributes should apply to both the systems themselves and of the content, data and processes they manage but there is no standard or benchmark on which we can measure the merit of a multimedia based content design. Technostics ignore the unique instructional capabilities of e-Learning by importing legacy materials from books or classroom manuals without employing engaging multimedia features as a result often e-Learning fails to live up to its potential, hence learning suffers.

- S. Sengupta is with the CSE and IT Department, Bengal Institute of Technology, Kolkata, India. E-mail: mesouvik@hotmail.com.
- S. Pal is with the CSE and IT Department, Bengal Institute of Technology, Kolkata, India. E-mail: saurabhpal2007@gmail.com.
- N. Banerjee is with the CSE and IT Department, Bengal Institute of Technology, Kolkata, India. E-mail: ban.nilanjan@gmail.com.

Our objective in this project is to find out an algorithm which will help us to compare between different e-Learning content with respect to compliance with LTSA layer 1 and SCORM. We intended to measure how a e-Learning content is using the different interaction components to interact with the user.

## 2 SCORM AND LTSA

The development of e-learning standards is a complicated, time-consuming, and challenging process. In the case of e-learning, groups like the Aviation Industry CBT Committee (AICC), IMS Global Learning Consortium (IMS), MedBiquitous, and the Schools Interoperability Framework (SIF) spend many months writing technical specifications that recommend how content should be developed, delivered, tracked, managed, stored, accessed, customized, etc.

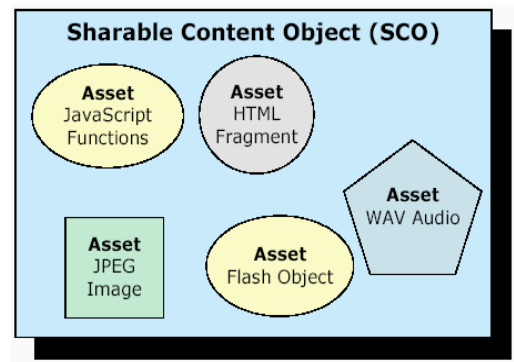

Fig 1

SCORM was developed by the US Advanced Distributed Learning (ADL) Initiative for e-learning solutions created by and for the US government. However, the worldwide adoption of SCORM by government, industry, and academic groups has led to its status, and prominence, as the *de facto standard* in e-learning.

A SCO is a collection of one or more Assets that represent a single learning resource that utilizes the SCORM Run-Time Environment (RTE) to communicate with Learning Management Systems (LMS). A SCO represents the lowest level of granularity of a learning resource that is tracked by an LMS. The figure 1 shows an example of a SCO composed of several assets, note that the technologies depicted are provided as examples. SCORM does not specify any particular schema, format, or template for the content itself. SCOs are intended to be subjectively small units, such that potential reuse across multiple learning contexts is feasible during content design, when determining the size of a SCO, thought should be given to the smallest logical size of content to be tracked [3]. Reuse requirements and other factors impact decisions about the size of SCOs for service or joint organizations. Other factors may include how much information is required to achieve a learning outcome and the point where a branching decision is required for sequencing. Our work in this paper is concerted on finding the java script element in a web page and specifically functions declared to load, run and terminate a particular service.

## LTSA

The five-layered LTSA standard specifies a high level architecture for information technology-supported learning, education, and training systems. This standard is pedagogically neutral, content-neutral, culturally neutral, and platform-neutral. The layer's specification is from abstraction towards implementation as one moves from Layer I to Layer V. The layer I is the highest level of abstraction where two entities Learner and Environment and their interactions are shown [1]. This IEEE Standard covers a wide range of systems, commonly known as learning technology, education and training technology, computer-based training, computer assisted instruction, intelligent tutoring, metadata, etc. LTSA provides a framework for understanding existing and future systems, promotes interoperability and portability by identifying critical system interfaces, and incorporates a technical horizon (applicability) of at least 5-10 years while remaining adaptable to new technologies and learning technology systems. This Standard is neither prescriptive nor exclusive[2].

## 3 RELATED WORKS

The LTSA Layer1 specifies abstract level architecture of learners interaction with the environment where as the SCORM specifies use of SCO with help of java script. There is a good number of work done already to create a platform where one can check the compliance of the said specifications. In [4] a research is done to find the problems with current eLearning platforms and recommend ways to improve the effectiveness and efficiency of eLearning platforms Some of these problems are indentified as : Text-based learning materials, Lack of rich content for good understanding, Insufficient interactivity or flexibility, Unstructured and isolated multimedia instructions. In this paper we propose to show an comparative results of the e-Learning sites to understand how competent they are to overcome such problem by satisfying LTSA layer 1 and SCORM specifications.

## 4 COMPLIANCE OF LTSA LAYER 1 AND SCORM FOR E-LEARNING SITES FROM AN INFORMATION TECHNOLOGY PERSPECTIVE

In LTSA Layer 1 all Learner Interaction is from the environment to the Learner Entity (inbound information) thus the "Learner Interactions" arrow points inward towards the Learner Entity [2]. From the specification of SCROM and LTSA layer 1 it is obvious that the impact of the interactive multimedia content can be a measure of quality for any eLearning content and the rapid development of eLearning raises a great demand on the use of a consistent set of standards [10]. Considering the fact that in a multimedia based e-Learning environment the learner generally interacts with images, sound and movie file, hyperlinks, text, download links, active contents [applet & activeX ] we count an plot the proportional frequency of the input e-Learning sites to form the bar chart. We also check for the java script function call to understand the scope of the content to be used as SCO. The portable multimedia components and hyper references in addition stimulate content exchange and composition between authors [9].

## 5 PROPOSED ALGORITHM

Step 1 : Set global variables
word_count=0,
image_count=0,
audio_count =0,
video_count = 0,
Active_content_count =0,
downLoadable_content_count = 0,
script_functions = 0,
form_control_count =0,
inbound_link_count = 0,
outbound_link_count = 0,
keyword_count = 0.

Set Visted_List as List

Step 2 : Select index/default/home web page Pi
Step 3 : call trace(Pi).
Step 4 : Calculate the percentage of images, audio, video, hyperlinks; downloadable, active contents usage with respect to the total element usage.
Step 5 : Plot the result in the bar chart.
Function trace(P)
{
Step 1: For ref in each hyper references of P
 {

```
     If( outbound(ref)==false)
         {
             Select web page Q corresponding to ref
             If Q not in Visted_List
{
         Call FindElements(Q)
                 Call trace(Q)
}
    }
  }
}

Function FindElements(P)
{
Set X, Y as String
For each character  Ci  in P
{
   If Ci is between '>' and '<' then
       X = X + Ci
   If Ci is between '<' and '>' then
       Y = Y + Ci

}

call WordCounter(X)
call ImageCounter(Y)
call ActiveContentCounter(Y)
call LinkCounter(Y)
call DownLoadableCounter(Y)
call ScriptCounter(X)
call AudioCounter(Y)
call VideoCounter(Y)
call FormElementCounter(Y)
Add P to Visited_List
}
Function WordCounter(X)
{
For each word  in  X;
{
      word_count = word_count + 1
}
}
Function ImageCounter(Y){
{
Srep 1: Search "src" pattern in X
Step 2: For each pattern found in X
       {
         Extract the file name associated with "src"
          Check the file extension for type ".bmp", ".jpg",
".gif"
If  file extension matches then
 image_count=image_count+1
}

}

Function ActiveContentCounter(Y)
{
Step 1: set count = 0;
Step 2 : Search "applet" pattern in Y
Step 3 : For each pattern found in Y
         {
              Count =count +1;
}
Step 4: active_ content_ count = active_ content_ count + (count/2);
Step 5: Search "src" pattern in Y
Step 6: For each pattern found in Y
         {
             Extract the file name associated with "src"
             Check the file extension for type ".swf"
              If  file extension matches then
             active_content_count =
                  active_content_count + 1;
}
}

Function LinkCounter(Y){
Step 1: Search "href" pattern in Y
Step 2: For each pattern found in Y
           {
 Set hyper_refernce = Extracted url (address and file name) associated with "href".
Search patterns "www", "http", ".html",".htm"
in hyper_refernce.
If ".htm" or ".html" pattern found in hyper_refernce
              {
                 Set flag = call outbound(hyper_refernce)
                  }
           Else if   "http" or "www" pattern found in hyper_refernce
                  {
     Set flag = call outbound(hyper_refernce)
                  }
 If (flag = = false) then
     {

 inbound_link_count = inbound_link_count +1
  }
 else
{
             outbound_link_count = outbound_link_count + 1
}
           }
Function DownLoadableCounter(Y){
{
Step 1: Search "href" pattern in Y
Step 2: For each pattern found in Y
         {
             Extract the file name associated with "href"
```

```
        Check the file extension for type ".doc", ".pdf", ".ppt"
If file extension matches then
 downLoadable_content_count= downLoadable_content_count +1
}
}
Function ScriptCounter(X)
{
Step 1: Set block_count = 0,flag = false.
Step 2: For each characters Ci in X
     {
       If (Ci == '{') then
       {
       Flag = false.
       block_count = block_count + 1
       }
       If (Ci == '}') then
       {
       Flag = true.
       block_count = block_count -1.
       }
}
       if(flag==true AND block==0)
       {
        script_functions = script_functions +1
        flag = false;
       }
     }
Function AudioCounter(X){
Srep 1: Search "src" pattern in Y
Step 2: For each pattern found in Y
       {
          Extract the file name associated with "src"
          Check the file extension for type ".wav", ".mp3"
If file extension matches then
 audio_count = audio_count +1
}

Function videoCounter(Y){
{
Step 1: Search "src" pattern in Y
Step 2: For each pattern found in Y
       {
          Extract the file name associated with "src"
          Check the file extension for type ".dat", ".avi"
If file extension matches then
 video_count = video_count +1
}

}
Function FormElementCounter(Y){
Step 1: Set count = 0.
Step 2: Search "input" pattern in Y
Step 3: For each pattern found in Y
         form_control_count = form_control_count + 1
Step 4: Search pattern "textarea", "select", "button" in Y.
Step 5: For each pattern found Y
         count = count + 1.
Step 6 : form_control_count = form_control_count + (count/2).

}
Function keyword_counter(X)
{
Step 1: For each word in X
   If word matches in lexicon domain then
       keyword_count = keyword_count + 1
           else
           {
String of consecutive 2 to 3 words are matched in domain lexicon

If word matches in lexicon domain
      keyword_count = keyword_count + 1

       }
}
```

## 6 OUTPUT

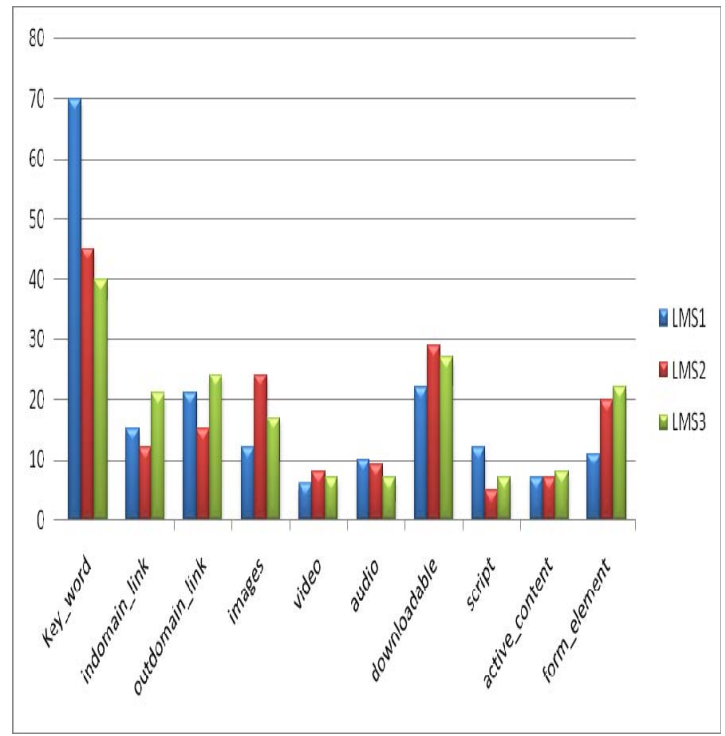

Fig 2

# 7 CONCLUSION

The work of this paper is divided in three major parts, the first part deals with studying and understanding what exactly LTSA Layer 1 and SCORM compliance mean and then we adopt the specifications in a e-Learning scenario and see that these sites are comparable with respect to specifications if we can have a chart showing its usage of multimedia components. Secondly we derived an algorithm which will calculate the proportional usage of different multimedia element in the web site by crawling through the pages. We have taken only html pages into account in this connection. Finally we draw a bar chart with the result obtained from the different e-Learning sites that we want to compare. The objective of this comparison is not for concluding which one is a better learning package but to understand which one is more close to LTSA and SCORM compliance so that we can step towards standardization.